\documentclass[trackchanges]{aastex701}

\shorttitle{The Local Sky as an Astronomy Laboratory}
\shortauthors{Hart}

\graphicspath{{./figures/}}

\begin{document}

\title{The Local Sky as an Introductory Solar System Astronomy Laboratory Using Smart-Telescopes}

\author[orcid=0000-0002-2945-5530,gname=Roger,sname=Hart]{Roger M. Hart}
\affiliation{Community College of Rhode Island, Flanagan Campus, 1762 Louisquisset Pike, Lincoln, RI 02865-4585, USA}
\email[show]{rmhart1@ccri.edu}
\correspondingauthor{Roger M. Hart}

\begin{abstract}

Portable smart telescopes can make introductory astronomy more observational, quantitative, and locally grounded by allowing students to acquire, share, and analyze astronomical images within a single class period. This article presents an eight-laboratory sequence for introductory Solar System astronomy using the Seestar S50 smart telescope and archived or student-collected images. The sequence begins with a galaxy-size investigation that introduces pixel scale, angular size, physical size, proportional reasoning, and measurement uncertainty before students apply the same image-to-evidence approach to Solar System objects. Subsequent activities examine the positions and apparent motion of Jupiter's Galilean moons, comparative planetology and telescope limitations, the apparent sizes of Mars, Jupiter, and Saturn, lunar phase and angular diameter, relative lunar surface history through crater-density comparisons, single-session solar activity, and multi-day sunspot tracking to estimate solar rotation. Each investigation follows a 5E-informed structure and produces a focused student product, such as a measurement table, graph, comparison chart, or claim-evidence-reasoning statement. Throughout the sequence, students must define transparent measurement rules, distinguish direct observations from interpretations, and evaluate how image scale, illumination, exposure, seeing, target visibility, and boundary selection constrain their conclusions. The activities are designed for high-school, dual-enrollment, community-college, and introductory university astronomy courses and may be implemented with live observations or prepared image sets when weather, scheduling, or local observing conditions prevent data collection.

\end{abstract}


\section{Introduction} \label{sec:introduction}

Astronomy is a natural setting for evidence-based instruction because every student already inhabits a sky. Introductory astronomy is commonly taught as a general education science course in three formats, either a one semester survey of all introductory astronomy including the solar system and stars and galaxies, or these topics separated into two separate semester length courses. In the solar system semester long course, here in as ASTR 1010, the challenge is to make that sky more than a list of objects or a backdrop for lecture. Place-based education and sense-of-place scholarship emphasize that setting can matter to learning, and observational astronomy is intrinsically local because horizon obstructions, sky glow, weather, season, Moon phase, and safe access determine what students can measure from a school courtyard, campus lawn, athletic field, or parking-lot edge \citep{ref01,ref02,ref03}. The 5E learning cycle provides a practical structure for this work. Students first encounter a phenomenon, then explore an image or live observation, explain the relevant concept, extend the method to a related case, and evaluate an evidence-based product \citep{ref04,ref05}. Smart and robotic telescopes make this structure more feasible because they reduce setup overhead and make images shareable and measurable during a single class period \citep{ref06,ref07}. The broader learning and standards literature supports this emphasis on disciplinary practices, models, evidence, and communication rather than vocabulary alone \citep{ref08,ref09,ref10,ref11}.

The teacher-facing literature already includes strong models for promoting scientific evaluation in astronomy, including evidence-comparison activities, Build a model evidence link (MEL) scaffolds, and research on plausibility judgments and scientific reasoning \citep{ref12,ref13,ref14,ref15,ref16}. Inexpensive image-based astronomy also has a long history, ranging from simple and low-cost astrophotography \citep{ref17,ref18,ref19} to smartphone astronomy, smartphone-based astrophotography, and smartphone analysis of planetary transits \citep{ref20,ref23,ref24}. Related classroom approaches use \textit{Tracker} for astronomical image analysis \citep{ref21}; planetary imaging and quantitative investigations involving astronomical distance, camera-based parallax, and the orbits of the Galilean moons \citep{ref22,ref26,ref27,ref28}; and astrophotography to engage non-STEM majors in scientific investigation \citep{ref25}. Observational research and night-laboratory experiences have likewise been designed to broaden student participation in authentic astronomy investigations \citep{ref29,ref30}. More recent work addresses electronic and robotic telescope laboratories and observing networks \citep{ref06,ref07,ref32,ref35}, portable smart telescopes and their use in education, citizen astronomy, and scientific observation \citep{ref31,ref36,ref37}, and course-based undergraduate astronomy research experiences and their effects on students \citep{ref33,ref34,ref77}. The present article builds on this prior work but narrows the emphasis to an ASTR 1010 sequence, focusing on solar system astronomy. Star clusters, wide binaries, and stellar-life-cycle classification remain useful in stars and galaxies course (ASTR 1020) activities \citep{ref58}. In this work, I show that portable smart telescopes allow students to collect and analyze astronomical images during class, making astronomy more observational and quantitative. This eight-lab Seestar S50 sequence explores galaxies, planets, moons, and the Sun and can be used in high-school or introductory college courses with live or prepared images that have a solar system focus.

\section{Classroom or instructional context} \label{sec:context}

These activities are intended for primary for an in-person or hybrid lecture online and laboratory in person, community college introductory solar system astronomy course but can also be deployed in high-school astronomy, dual-enrollment Earth and space science, and introductory university solar-system courses. The ideal time design is to fit within an hour meeting time. The two lunar modules can be used as one-period archived-data investigations or combined into a longer lunar project across a lunation. A practical class begins with a short Engage prompt, uses live observing when conditions cooperate, shifts quickly to original or backup images, and ends with a compact product such as a table, graph, comparison chart, or claim-evidence-reasoning paragraph. The ASTR 1010 course these were designed for meets for 5 hours per week and has approximately 15 hours of coursework outside of the course per week. 

One telescope can support a full class if roles are rotated deliberately, however using four students to one telescopes works best with our enrollment in our community college introductory solar system astronomy course. Useful roles include telescope operator, recorder, measurer, and skeptic-checker. The skeptic-checker is useful because there are often observational constraints and they can ask questions like, whether a galaxy edge rule is consistent, whether a point near Jupiter is a moon or field star, whether a planet boundary can be measured, whether a crater count uses the same area, and whether a sunspot conclusion goes beyond the image. Roles are coached early in the semester but the instructor fades group support and logistics as time progresses. Prepared images are not a lesser substitute for live data; weather, target visibility, daylight scheduling, and local obstructions are part of observational astronomy, however we try to substitute activities to ensure authentic observations. Our observations are in Lincoln, Rhode Island, USA with moderately heavy light pollution. Students need only modest mathematics, which mirrors our course: proportional reasoning, pixel measurements, units, simple ratios, and uncertainty language. This fits within the astronomy course math requirements. Screen sharing, high-contrast packets, verbal figure descriptions, role rotation, and analysis from prepared images help every student participate in the image-to-claim process. These smart-telescope laboratories can be implemented as complete, stand-alone investigations or used selectively to supplement existing curricula, including activities developed by the NASA Heliophysics Education Activation Team (NASA HEAT). Because smart-telescope observations connect astronomical imaging and measurement with concepts in physics, Earth science, and space science, the laboratories can support the integration of Earth and space science contexts into physics instruction \citep{ref78}. Individual investigations may also be paired with established space-science learning sequences addressing electromagnetic radiation, electromagnetic fields, and related heliophysics concepts \citep{ref79}. The solar modules can extend eclipse-centered undergraduate instruction by linking observations of the Sun to broader course content and major public observing events \citep{ref80}. For example, although a smart telescope does not directly measure magnetic fields, its planetary observations can provide an observational anchor for complementary instruction on planetary magnetism in Solar System astronomy \citep{ref81}. This flexibility allows instructors to select observations and analyses that fit their course content, available class time, local observing conditions, and access to complementary curricular resources.

\section{Physics background} \label{sec:physics}

The activities were developed around the Seestar S50 because it combines a small telescope, camera, alt-azimuth mount, focusing, app control, and image export in a portable system. The classroom working values used in the source materials are a 50 mm aperture, 250 mm focal length, native 1920 x 1080 images, a field of view of about 1.29 deg x 0.73 deg, and a working plate scale near 2.4-2.42 arcsec/pixel \citep{ref38,ref39,ref40,ref41,ref42}. The plate scale should be treated as an instructional approximation unless the class calibrates it with a known star field.

For an image length $N_{\rm px}$ measured in pixels, the angular size or separation is

\begin{equation}
\theta = N_{\rm px}s, \qquad s \approx 2.42~\mathrm{arcsec~pixel^{-1}},
\label{eq:angular}
\end{equation}

Here $\theta$ is in arcseconds and $s$ is the plate scale. Physical size follows from the small-angle relation

\begin{equation}
L = \frac{d\theta}{206{,}265},
\label{eq:physical}
\end{equation}

where $L$ and $d$ have the same distance unit and $\theta$ is in arcseconds. Equations~(\ref{eq:angular}) and~(\ref{eq:physical}) support the first galaxy-scale lab and the lunar angular-diameter work. For lunar surface comparisons, an introductory crater-density measure can be defined as

\begin{equation}
\rho_{c} = \frac{N_{\rm craters}}{A_{\rm region}},
\label{eq:craterdensity}
\end{equation}

where $A_{\rm region}$ is a standardized area in pixels squared or a normalized region drawn on the same image. The relative terrain comparison is

\begin{equation}
R = \frac{\rho_{\rm highland}}{\rho_{\rm mare}},
\label{eq:terrainratio}
\end{equation}

For the solar-activity snapshot, students can record sunspot count or normalized spot position,

\begin{equation}
u = \frac{x}{R_{\rm disk}},
\label{eq:normalizedspot}
\end{equation}

where $x$ is the measured spot position from disk center and $R_{\rm disk}$ is the solar-disk radius in pixels. If a multi-day sequence of the same spot group is supplied, the simple solar-rotation estimate is

\begin{equation}
P_{\rm Sun} \approx \frac{2D_{\rm disk}}{v},
\label{eq:solarrotation}
\end{equation}

where $D_{\rm disk}$ is the solar-disk diameter in pixels and $v$ is the measured sunspot speed in pixels/day across the visible disk. This introductory estimate treats the apparent sunspot motion across the disk as uniform and should be interpreted as an approximate synodic rotation period. The most important lesson across these equations is that each number depends on a measurement rule: galaxy edges are threshold choices, lunar crater counts depend on illumination and region definition, planet diameters can be limited by saturation, and sunspots are changing tracers rather than fixed marks on a solid surface \citep{ref41,ref43,ref44,ref45,ref46,ref47,ref48}.

\section{Description of the Laboratories} \label{sec:activities}

\begin{figure*}[ht!]
\centering
\includegraphics[width=\textwidth]{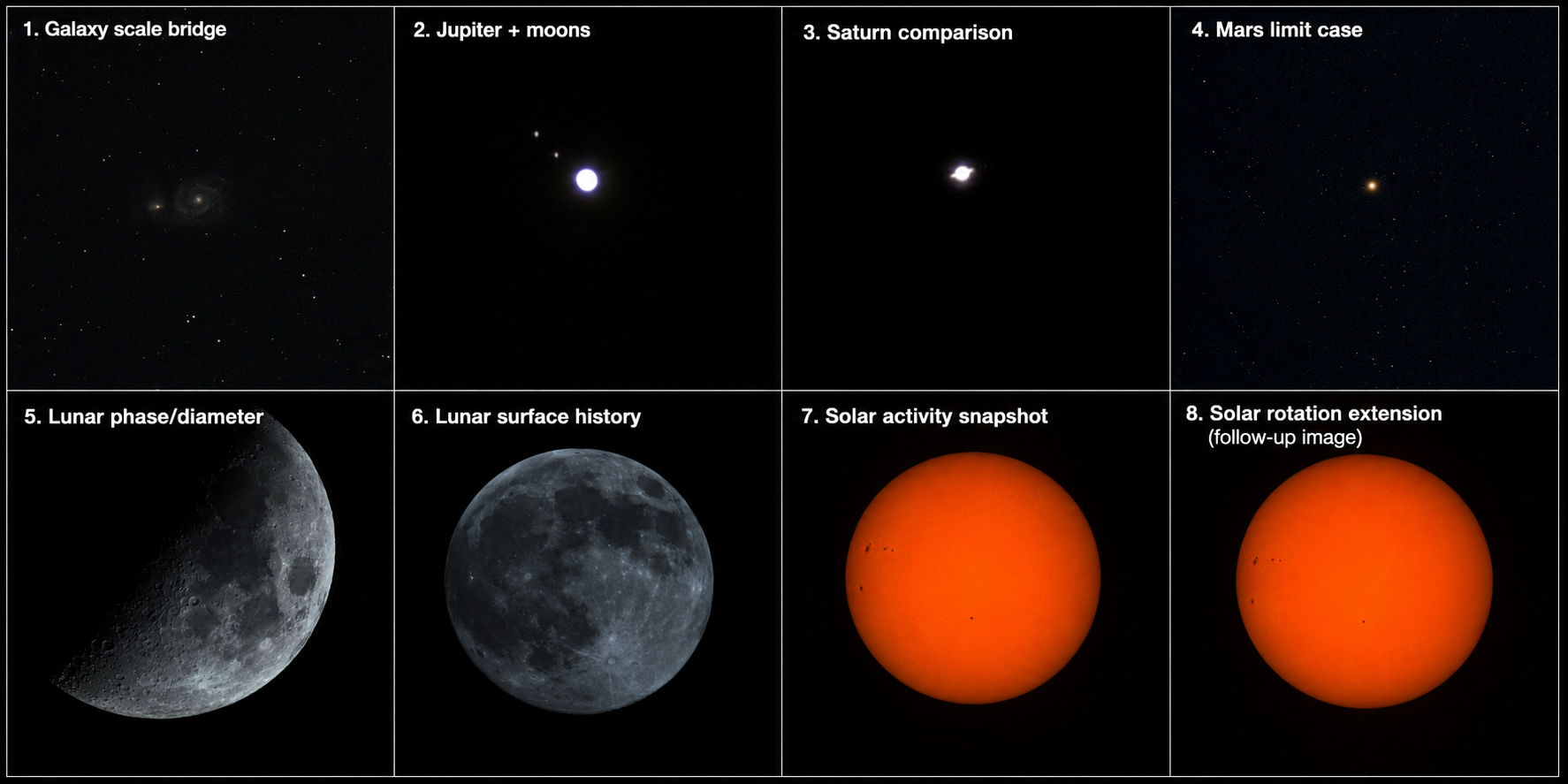}
\caption{Selected image set for the revised ASTR 1010 sequence. The panels show a galaxy target used as the first scale bridge; Jupiter with nearby point sources; Saturn; Mars as a small apparent-size and image-limit case; a lunar phase image for diameter and terminator work; a fuller lunar image for maria/highlands and crater-density comparison; a solar image for sunspot identification; and the same solar context as a bridge to multi-day rotation work. The figure is a teaching composite, not a calibrated science product. Students should measure from original native image files rather than from reduced manuscript panels.}
\label{fig:sequence}
\end{figure*}

\begin{table*}[ht!]
\centering
\caption{Revised ASTR 1010 smart-telescope investigation sequence.}
\label{tab:sequence}
\renewcommand{\arraystretch}{1.15}
\scriptsize
\begin{tabular}{c l l l l}
\hline
\textbf{Lab} & \parbox[t]{0.22\textwidth}{\raggedright \textbf{Module}} & \parbox[t]{0.19\textwidth}{\raggedright \textbf{Primary idea}} & \parbox[t]{0.22\textwidth}{\raggedright \textbf{Student product}} & \parbox[t]{0.16\textwidth}{\raggedright \textbf{Best fit}} \\
\hline
1 & \parbox[t]{0.22\textwidth}{\raggedright Galaxy size in smart-telescope images} & \parbox[t]{0.19\textwidth}{\raggedright Angular size, physical size, and scale} & \parbox[t]{0.22\textwidth}{\raggedright Major/minor-axis measurements and size estimate} & \parbox[t]{0.16\textwidth}{\raggedright ASTR 1010 scale bridge} \\
2 & \parbox[t]{0.22\textwidth}{\raggedright Jupiter and the Galilean moons} & \parbox[t]{0.19\textwidth}{\raggedright Orbital motion in a planetary system} & \parbox[t]{0.22\textwidth}{\raggedright Moon-position table and offset plot} & \parbox[t]{0.16\textwidth}{\raggedright ASTR 1010} \\
3 & \parbox[t]{0.22\textwidth}{\raggedright Comparative planetology and telescope limits} & \parbox[t]{0.19\textwidth}{\raggedright Jupiter/Saturn comparison and image limits} & \parbox[t]{0.22\textwidth}{\raggedright Observation/inference comparison chart} & \parbox[t]{0.16\textwidth}{\raggedright ASTR 1010} \\
4 & \parbox[t]{0.22\textwidth}{\raggedright Planetary apparent size: Mars, Jupiter, and Saturn} & \parbox[t]{0.19\textwidth}{\raggedright Apparent diameter, distance, seeing, and saturation} & \parbox[t]{0.22\textwidth}{\raggedright Apparent-size/limits comparison} & \parbox[t]{0.16\textwidth}{\raggedright ASTR 1010} \\
5 & \parbox[t]{0.22\textwidth}{\raggedright Lunar Science I: phase, diameter, and terminator visibility} & \parbox[t]{0.19\textwidth}{\raggedright Lunar phase, angular size, crater visibility} & \parbox[t]{0.22\textwidth}{\raggedright Phase/diameter/crater record} & \parbox[t]{0.16\textwidth}{\raggedright ASTR 1010} \\
6 & \parbox[t]{0.22\textwidth}{\raggedright Lunar Science II: surface history from crater density} & \parbox[t]{0.19\textwidth}{\raggedright Maria/highlands and relative surface age} & \parbox[t]{0.22\textwidth}{\raggedright Standardized crater-density comparison} & \parbox[t]{0.16\textwidth}{\raggedright ASTR 1010; new lunar module} \\
7 & \parbox[t]{0.22\textwidth}{\raggedright Solar Activity Snapshot} & \parbox[t]{0.19\textwidth}{\raggedright Sunspot identification, position, and image evidence} & \parbox[t]{0.22\textwidth}{\raggedright Sunspot group count/position table and CER claim} & \parbox[t]{0.16\textwidth}{\raggedright ASTR 1010; new collected-image module} \\
8 & \parbox[t]{0.22\textwidth}{\raggedright Sunspot tracking and solar rotation} & \parbox[t]{0.19\textwidth}{\raggedright Rotation inferred from repeated tracers} & \parbox[t]{0.22\textwidth}{\raggedright Sunspot-position table and period estimate} & \parbox[t]{0.16\textwidth}{\raggedright ASTR 1010 extension} \\
\hline
\end{tabular}
\end{table*}

Table~\ref{tab:sequence} summarizes the ASTR 1010 main sequence of activities. These laboratories an be found at https://www.ccri.edu/faculty-staff/rmhart1/. In Lab 1, Galaxy Size in Smart-Telescope Images, students measure a galaxy's major and minor axes in pixels, convert the values to angular size, and use an instructor-supplied distance to estimate physical size. The major point in this first part of the lab collection is scale: apparent size and physical size are different, and a visible galaxy boundary is an edge rule shaped by exposure depth, surface brightness, background sky, and processing. I suggest M51 and M81 being strong first targets, whereas M31 remains useful as a challenge because a single frame often captures only the bright central region \citep{ref26,ref41,ref46}. For Lab 2, Jupiter and the Galilean Moons, students infer orbital motion from repeated image measurements rather than label recognition. Students mark Jupiter's center, measure horizontal offsets of visible moon candidates, convert pixels to angular separation, and decide which points move with Jupiter rather than behave like background stars. The included Jupiter images support identification and offset practice; a longer same-night sequence is needed for ranking moon speeds with confidence \citep{ref49,ref50,ref51,ref52,ref53,ref54,ref55}. In Lab 3, Comparative Planetology and Telescope Limits, students compare Jupiter and Saturn by separating direct image evidence from prior knowledge. Students record rings, glare, point sources, disk shape, saturation, approximate apparent size, and features hidden by seeing or exposure. This will not be a perfect planetary portrait but a comparison of physical differences and observing limitations \citep{ref56,ref57}. Because the images do not directly reveal planetary magnetic fields, the module can be paired with a complementary lesson on planetary magnetism \citep{ref81}. In Lab 4, Planetary Apparent Size: Mars, Jupiter, and Saturn, this shows us the telescope used in ASTR 1010 image-limits activity using the collected Mars, Jupiter, and Saturn observations. Mars is especially valuable because it can appear almost star-like or saturated in a small smart-telescope image. Students learn that angular diameter, distance, exposure, seeing, and boundary definition all affect whether an apparent-size measurement is defensible. They should not treat images from different dates and exposures as a simple physical size comparison without caveats \citep{ref41}. In Lab 5, Lunar Science I, the premise is phase-and-diameter emphasis. Students compare phase, terminator placement, whole-disk diameter, crater visibility, and observation conditions. Full or nearly full images are better for disk diameter and maria/highland mapping, while first- and last-quarter images are better for relief and crater shadows. The activity can be one period with archived images or a multi-session thread across a lunation \citep{ref13,ref41,ref46}.

Lab 6, Lunar Science II: Surface History from Crater Density, has students placing equal-area boxes on a mare region and a highland region, count craters above a class-defined size threshold, compute a crater-density ratio, and write a claim about relative surface history. The key teaching point is that crater counts are evidence only when the region, threshold, phase, and illumination are stated. The lab also makes a useful bridge to impact history and planetary surface ages without requiring professional crater-statistics methods. This pairs well with a hands on impact crater lab and online lab that students complete. 

Lab 7, Solar Activity Snapshot, students work through sunspot identification, position practice, and discussions of projection and image quality, even though they do not by themselves form a multi-day rotation sequence, students in other classes help complete the dataset. Students mark the solar limb or usable visible disk, count spot groups, record normalized positions, and explain what a single solar session can and cannot prove \citep{ref44}.

Lab 8, Sunspot Tracking and Solar Rotation, remains as an extension or second solar period if time allows and it is a class that meets multiple times in a week. It uses the same measurement-first logic but requires a dated multi-day sequence of the same sunspot group. Students measure disk diameter and spot position over several days, estimate a rotation period, and explain why projection, differential rotation, and sunspot evolution limit precision \citep{ref44,ref59}.

\begin{figure*}[ht!]
\centering
\includegraphics[width=\textwidth]{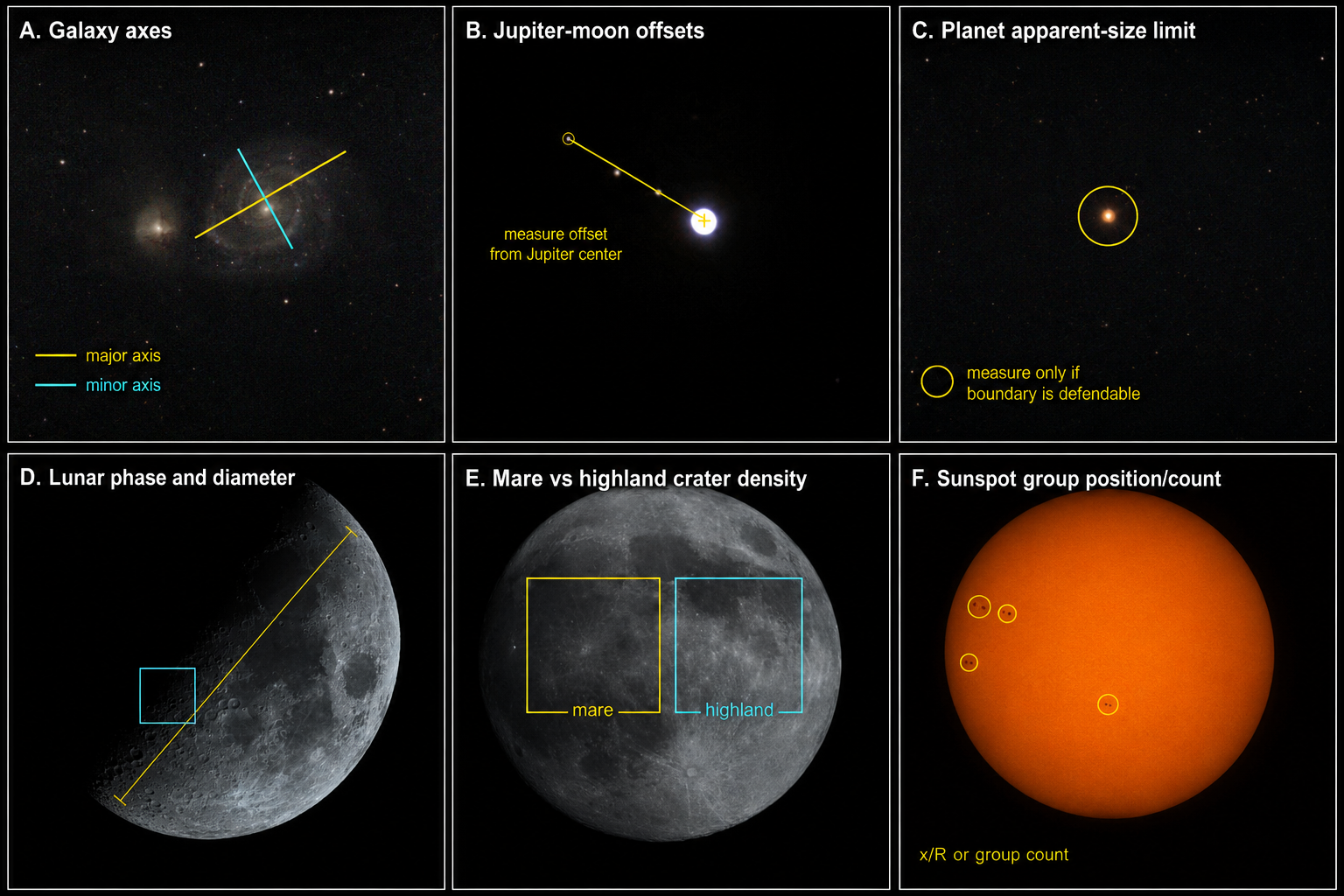}
\caption{From image to measurement rule. Panel A shows a galaxy major/minor-axis rule. Panel B shows Jupiter-moon offset measurement from a chosen planet center. Panel C shows why a planet boundary may be difficult to defend in a saturated or seeing-limited image. Panel D shows a lunar diameter chord and a crater-count box. Panel E shows separate mare and highland boxes for the new crater-density lab. Panel F shows a sunspot group-count and normalized-position rule. The overlays make measurement choices visible; they do not assert unique correct values. Students should record the center, edge, limb, threshold, region, or coordinate rule used and state the dominant uncertainty.}
\label{fig:measurementrules}
\end{figure*}

\section{Methods and Results}
\label{sec:methods}

Across the sequence, the student products, a galaxy-size estimate, moon-offset table, planet evidence chart, apparent-size comparison, lunar phase/diameter record, lunar crater-density comparison, sunspot snapshot table, or solar-rotation estimate are reasonable given the time constraints of the course: 5 contact hours and 15 hours outside of class per week. They are being asked to decide what quantity can be measured, use a transparent rule, and connect the result to a claim.

I recommend that teachers should supply target lists, lunar phase windows, galaxy distances, solar safety instructions, and backup images before class. Distance lookup during a lab can distract from the intended measurement reasoning. Students should measure from original or native exported files when pixel scale matters because screenshots, embedded document previews, and resized images can change the effective scale.

Repeated measurements by two students are usually more valuable than a long derivation I think because they surface exactly the disagreements that matter within this lab set: galaxy edge choice, Jupiter-center choice, planet boundary choice, limb definition, crater-count threshold, and whether a sunspot group is counted as one feature or several.

Students immediately encounter the difference between apparent size, angular size, physical size, and measurement threshold. The next modules then apply the same habits to the Solar System: Jupiter's moons, planet comparisons, Mars/Jupiter/Saturn apparent size, the Moon, and the Sun.

A typical one lab meeting for implementation uses 5-10 minutes for an Engage prompt, 15-25 minutes for live image acquisition or archived-image inspection, 15-20 minutes for measurement, and 5-10 minutes for explanation and evaluation. The two lunar labs can be run separately: Lab 5 for phase/diameter/terminator reasoning and Lab 6 for surface-history crater counts. They can also be combined into a longer lunar unit if schedule and weather permit.

The solar modules should be sequenced carefully. Solar Activity Snapshot can be completed with the collected single-session images. Solar Rotation requires additional multi-day images of the same spot group. Treating those as two related but different labs makes the limitation clear rather than hiding it. Public events such as eclipse-viewing days can also connect undergraduate course content, which I have employed in the past, safe solar observing, and community outreach \citep{ref80}; local-news references and permissions should still be verified before publication \citep{ref76}.

The expected outcomes are teacher-facing rather than empirical claims. After the sequence, students should be able to define a measurement rule, distinguish observation from inference, convert pixels to angular quantities, use the small-angle relation appropriately, and name at least one limitation that changes the strength of a claim. Predictable difficulties become useful teaching targets and you may find that students may treat a galaxy's visible edge as exact, identify field stars as moons, infer too much from a single frame, treat Mars brightness as a direct size measurement, assume full moon is best for all lunar questions, count craters without a shared threshold, treat sunspots as permanent marks on a solid surface, or expect a one-session solar set to yield a rotation period. I think one good aspect of these labs is what does the data collection and interpretation mean with respect to my question. 

\section{Discussion} \label{sec:discussion}

The main genesis of this laboratory collection was to leverage students performing authentic astronomical observations with other activities, so that combined there is high engagement with the course, greater student success, and more students having an appreciation of astronomy and understanding the process of science. For example, the Solar Activity Snapshot lab is deliberately different from Solar Rotation. It is justified by the collected data: single-session solar images can support sunspot identification and position practice, while a rotation-period estimate requires a multi-day sequence. This distinction is pedagogically valuable because it shows students that evidence can be useful without supporting every desired conclusion. The second lunar lab likewise separates two questions that are often collapsed. Lunar Science I asks how phase and geometry affect what we see. Lunar Science II asks how crater counts can serve as evidence for relative surface history. The same images can support both, but the measurement rules differ. This helps students understand that an astronomical image is not one piece of evidence; it is a source from which different evidence can be extracted under different rules.

The same design logic can support future extensions such as Io/Roemer timing, Stellarium reconstructions, asteroid occultations, parallax and astronomical-unit work, simple spectrography, variable-star analysis, citizen-science networks, heliophysics, low-cost magnetometers, and GPS-based Earth-measurement comparisons \citep{ref60,ref61,ref62,ref63,ref64,ref65,ref66,ref67,ref68,ref69,ref70,ref71,ref72,ref73,ref74,ref75}. Several cautions should remain explicit, such as observational and laboratory safety. For example, seasonal visibility and local obstructions determine which targets are available. Galaxy-size estimates depend strongly on the edge rule and on supplied distances. Bright planets may saturate or blur in deep-sky-oriented images. Apparent-size comparisons across different dates and exposures require caveats. Lunar crater counts depend on phase, illumination, image scale, and threshold choice. Daylight-blue lunar images can be useful precisely because they show how contrast and limb definition affect measurement quality. Any Solar work requires the correct solar filter, instructor verification before pointing, and continuous supervision. Official Seestar documentation warns against observing the Sun directly without the solar filter \citep{ref40}.

\begin{table*}[ht!]
\centering
\caption{Dominant uncertainty and practical caution by ASTR 1010 module.}
\label{tab:uncertainty}
\renewcommand{\arraystretch}{1.15}
\scriptsize
\begin{tabular}{l l l}
\hline
\parbox[t]{0.19\textwidth}{\raggedright \textbf{Module}} & \parbox[t]{0.31\textwidth}{\raggedright \textbf{Dominant uncertainty or practical limit}} & \parbox[t]{0.40\textwidth}{\raggedright \textbf{Teacher-facing caution}} \\
\hline
\parbox[t]{0.19\textwidth}{\raggedright Galaxy size} & \parbox[t]{0.31\textwidth}{\raggedright Boundary choice, surface brightness, supplied distance} & \parbox[t]{0.40\textwidth}{\raggedright Use one or two targets; require an edge rule before students calculate physical size.} \\
\parbox[t]{0.19\textwidth}{\raggedright Jupiter moons} & \parbox[t]{0.31\textwidth}{\raggedright Glare, hidden moons, short time baseline} & \parbox[t]{0.40\textwidth}{\raggedright A missing moon is a scientific limitation, not a failed observation.} \\
\parbox[t]{0.19\textwidth}{\raggedright Comparative planetology} & \parbox[t]{0.31\textwidth}{\raggedright Saturation, seeing, exposure choice} & \parbox[t]{0.40\textwidth}{\raggedright Separate direct image evidence from prior knowledge.} \\
\parbox[t]{0.19\textwidth}{\raggedright Planetary apparent size} & \parbox[t]{0.31\textwidth}{\raggedright Different dates/exposures, small disks, saturation} & \parbox[t]{0.40\textwidth}{\raggedright Do not treat brightness or apparent diameter as simple physical size evidence without caveats.} \\
\parbox[t]{0.19\textwidth}{\raggedright Lunar Science I} & \parbox[t]{0.31\textwidth}{\raggedright Phase timing, limb/terminator visibility, orientation} & \parbox[t]{0.40\textwidth}{\raggedright Use the phase that matches the question.} \\
\parbox[t]{0.19\textwidth}{\raggedright Lunar Science II} & \parbox[t]{0.31\textwidth}{\raggedright Crater threshold, terrain choice, illumination} & \parbox[t]{0.40\textwidth}{\raggedright Use equal-area regions and report threshold rules.} \\
\parbox[t]{0.19\textwidth}{\raggedright Solar Activity Snapshot} & \parbox[t]{0.31\textwidth}{\raggedright Single-session data, limb/projection effects, spot grouping} & \parbox[t]{0.40\textwidth}{\raggedright Use for identification and position practice, not rotation period by itself.} \\
\parbox[t]{0.19\textwidth}{\raggedright Solar rotation} & \parbox[t]{0.31\textwidth}{\raggedright Projection effects, differential rotation, sunspot evolution} & \parbox[t]{0.40\textwidth}{\raggedright Solar filter installation and instructor verification are mandatory; add multi-day images.} \\
\hline
\end{tabular}
\end{table*}

\section{Conclusion} \label{sec:conclusion}
A small, portable smart telescope can make introductory Solar System astronomy more observational, quantitative, and evidence-centered. In this eight-laboratory sequence, students begin with galaxy-size measurements that establish pixel scale, angular size, physical size, proportional reasoning, and uncertainty before applying the same image-to-evidence approach to Jupiter’s moons, planetary comparisons and telescope limitations, lunar phase and surface history, solar activity, and solar rotation. Through the 5E-informed investigations, students acquire or analyze images, define transparent measurement rules, produce focused data products, distinguish observations from interpretations, and evaluate the limitations imposed by image scale, illumination, exposure, seeing, visibility, and boundary selection. Because the activities can be completed with either live observations or prepared image sets, they remain practical when weather, scheduling, or observing conditions prevent data collection. Together, these investigations demonstrate that a compact smart telescope and a local observing site can transform the familiar sky into a measurable laboratory in which students construct and evaluate scientific claims from evidence.

\section{Acknowledgments} \label{sec:acknowledgments}
The author acknowledges the \textit{ESS Scaffolded Inquiry Labs} series developed by Smay, Kortz, Hart, and Rogers. This 5E-based laboratory collection is designed specifically to supplement the laboratories in that series for use in ASTR 1010. This material is based upon work supported by the National Science Foundation under Grant No. 2514197. Any opinions, findings, conclusions, or recommendations expressed in this material are those of the authors and do not necessarily reflect the views of the National Science Foundation.

\bibliography{hart_seestar_references}{}
\bibliographystyle{aasjournalv7}

\end{document}